\documentclass[10pt]{iopart}
\usepackage{iopams}

\expandafter\let\csname equation*\endcsname=\relax
\expandafter\let\csname endequation*\endcsname=\relax
\usepackage{amsmath}
\usepackage{hyperref}

\usepackage{amssymb}
\usepackage{amsfonts,latexsym}
\usepackage{bm}
\usepackage[mathcal]{euscript}
\usepackage{graphicx}
\usepackage{epsfig}
\usepackage{color}

\newcommand{\be}{\begin{equation}}
\newcommand{\ee}{\end{equation}}

\newcommand{\mC}{\mathcal{C}}

\begin{document}

\title[Hidden oscillations in a disordered mean-field spin model]{Hidden collective oscillations in a disordered mean-field spin model with non-reciprocal interactions}


\date{\today}
\author{Laura Guislain and Eric Bertin}

\address{Univ.~Grenoble Alpes, CNRS, LIPhy, F-38000 Grenoble, France}


\begin{abstract}
We study the effect of introducing separable quenched disorder on a non-equilibrium mean-field spin model exhibiting a phase transition to an oscillating state in the absence of disorder, due to non-reciprocal interactions. In the disordered model, the magnetisation and its time derivative no longer carry the signature of the phase transition to an oscillating state. 
However, thanks to the separable (Mattis-type) form of the disorder, the presence of oscillations can be revealed by introducing a specific, disorder-dependent observable.
We also introduce generalised linear and non-linear susceptibilities associated either with the magnetisation or with its time derivative.
While linear susceptibilities show no sign of a phase transition, the third-order susceptibilities present a clear signature of the onset of an oscillating phase.
In addition, we show that the overlap distribution also provides evidence for the presence of oscillations, without explicit knowledge of the disorder.
\end{abstract}


\section{Introduction}

The onset of spontaneous collective oscillations in systems driven far from equilibrium is an intriguing phenomenon, observed for instance in biochemical clocks \cite{Cao_free_energy2015,nguyen_phase_2018,Aufinger_complex2022}, interacting active particles \cite{saha_scalar_2020,you_nonreciprocity_2020}, assemblies of 
biological cells \cite{Kamino_fold2017,Wang_emergence2019}, and population dynamics \cite{andrae_entropy_2010,Duan_Hopf2019}.
Such oscillations have also been reported in more abstract models like socio-economic models \cite{Gualdi2015,yi_symmetry_2015} or non-equilibrium Ising-type spin models \cite{collet_macroscopic_2014,collet_rhythmic_2016,collet_effects_2019,de_martino_oscillations_2019,martino_feedback2019,daipra_oscillatory_2020,guislain_nonequil2023,guislain_discontinuous2024},
for which the role of non-reciprocal interactions has been recently outlined \cite{avni_non-reciprocal2023}. In most of the above systems, the onset of spontaneous oscillations is clearly an emerging phenomenon that differs from the well-studied phenomenon of oscillator synchronization \cite{acebron_kuramoto_2005,risler_universal_2004}, since basic units like spins do not oscillate individually.
In particular, the collective order parameter does not simply reflect a breaking of the symmetry of individual units (e.g., spin-reversal symmetry) \cite{guislain_nonequil2023}.

In the mean-field case, the transition to spontaneous temporal oscillations can be described in the infinite system size limit as a Hopf bifurcation \cite{crawford_introduction_1991}, since the dynamics becomes deterministic in this limit.
However, fluctuations play a significant role in mesoscopic systems such as biochemical clocks \cite{Fei_design2018}, for which the coherence time of oscillations becomes finite \cite{gaspard_correlation_2002,barato_cost_2016,barato_coherence_2017,oberreiter_universal_2022,remlein_coherence_2022}.
In such situations, the noisy transition to spontaneous temporal oscillations may be described as a stochastic Hopf bifurcation \cite{Sagues2007,Xu_Langevin2020}, or in the framework of a generalised Landau theory \cite{guislain_nonequil2023,guislain_discontinuous2024}.

In this paper, we investigate in the framework of a Mattis-like driven disordered spin model how the presence of disorder in mean-field spin models with non-reciprocal interactions may lead to hidden oscillations, i.e., oscillations that are only visible using specific disordered-dependent observables (no sign of oscillations being visible on the magnetisation).
Interestingly, we observe that oscillations can be detected even with a partial knowledge of the microscopic disorder.
In addition, we found that the overlap distribution of spin configurations exhibits a non-trivial shape, with in particular a finite support, in qualitative analogy with the results found in a spin-glass phase in the case of a continuous replica symmetry breaking.
We show that in spite of the presence of disorder in the model, the non-trivial overlap distribution results from hidden oscillations and not from a spin-glass behaviour.

\section{Model and hidden phase transition}

\subsection{Definition of the model}

We introduce a disordered generalization of the kinetic mean-field Ising model with ferromagnetic interactions introduced in \cite{guislain_nonequil2023,guislain_discontinuous2024}
--see also \cite{collet_macroscopic_2014,collet_rhythmic_2016,collet_effects_2019,de_martino_oscillations_2019,martino_feedback2019,daipra_oscillatory_2020,avni_non-reciprocal2023} for related spin models. 
The model involves $2 N$ microscopic variables: $N$ spins $s_i=\pm 1$ and $N$ fields $h_i=\pm 1$.
The stochastic dynamics consists in randomly flipping a single spin $s_i$ or a single field $h_i$. We note $W_k^{s_i}$ the transition rates when flipping a spin $s_i=\pm1$ for $k=1$ or a field $h_i=\pm1$ for $k=2$. 
The transition rates when flipping a spin or field are taken as 
\be \label{eq:transition:rates} W_k^{s_i}=\left[1+\exp \left(\beta \Delta_i E_{k}\right)\right]^{-1},\ee 
with $\beta=T^{-1}$ the inverse temperature and $\Delta_i E_{k}$ the variation of $E_{k}$ when flipping a spin $s_i$ (for $k=1$) or a field $h_i$ (for $k=2$). 
We consider one of the simplest form of disorder by introducing separable quenched disorder, for which coupling constants factorize as a product of site-dependent quenched random variables.
We thus define the disorder by introducing on each site $i$ the quenched random variables $\{\epsilon_i\}_{1\leq i\leq N}$ which are independent and identically distributed random variables such that $\epsilon_i=\pm1$ with equal probability.
We denote as $\overline{x}$ the average over disorder of an arbitrary quantity $x$. This type of disorder was first introduced in \cite{mattis1976} for spin models exhibiting a paramagnetic to ferromagnetic phase
transition.\footnote{See also \cite{camilli-inference2022} for a recent contribution using a Mattis-like model in an inference context.}

The interaction energies between the spin and the field variables are given by 
\be \label{eq:energy:Es}E_s=-N^{-1}\sum_{i,j}\epsilon_i\epsilon_j\left(\frac{J_1}{2} s_i s_j+\frac{J_2}{2} h_ih_j+s_ih_j\right)\ee 
and 
\be \label{eq:energy:Eh}E_h=E_s+\mu N^{-1}\sum_{i,j} \epsilon_i\epsilon_js_ih_j.\ee
The particular case where disorder is absent, i.e., all $\epsilon_i=+1$, has been studied in \cite{guislain_nonequil2023,guislain_discontinuous2024}. Detailed balance is broken as soon as $\mu \neq 0$. Depending on $(T, \mu)$ values, the non-disordered model exhibits a paramagnetic (high $T$), ferromagnetic (low $T$, low $\mu$) or oscillating (low $T$, high $\mu$) behaviour. The three phases meet at a tricritical point $(T_c, \mu_c)$ with $T_c=(J_1+J_2)/2$ and $\mu_c=1+(J_1-J_2)^2/4$. 
From this mean-field model exhibiting a phase transition to an oscillating behaviour, we aim at understanding the effect of a quenched separable disorder, introduced through the variables $\epsilon_i$,
on a spontaneously oscillating system.

\subsection{Hidden collective oscillations}
\label{sec:hidden:oscill}

\begin{figure}[t]
    \centering
    \includegraphics{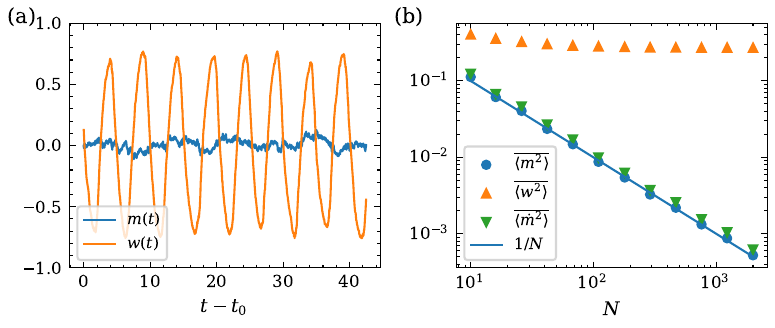}
    \caption{Stochastic simulations of the disordered mean-field spin model with non-reciprocal interactions. (a) Magnetisation $m(t)$ and disorder-dependent observable $w(t)$ for $N=470$. (b) $\overline{\langle m^2\rangle}$, $\langle w^2\rangle$ and $\overline{\langle \dot{m}^2\rangle}$ versus $N$. The plain line corresponds to the analytical prediction. Parameters for both panels: $T/T_c=0.6$, $\mu=1.4$, $J_1=0.6$, $J_2=0.4$.}
    \label{fig:m:t:m2:md2}
\end{figure}
On general grounds, phase transitions are characterized by order parameters associated with broken symmetries at the transition.  
In spin models without disorder, the magnetisation defined as 
$m=N^{-1}\sum_{i=1}^N s_i,$ is an order parameter associated with spin-reversal symmetry, which is only present in the paramagnetic phase. 
In the model without disorder, for low temperature and high $\mu$, one observes temporal oscillations of the magnetisation $m(t)$ with a non-zero amplitude $\langle m^2\rangle$.
For the model with separable disorder, we compute these quantities using stochastic simulations. In Fig.~\ref{fig:m:t:m2:md2}(a), we plot an example of trajectory of the magnetisation $m(t)$,
as well as the amplitude $\overline{\langle m^2\rangle}$ for different system sizes $N$ for a fixed temperature ($T<T_c$) and fixed $\mu$. The magnetisation $m(t)$ shows no sign of periodic oscillations, but only small random fluctuations. We determine the amplitude of these fluctuations by measuring $\overline{\langle m^2\rangle}$ as a function of system size $N$, as shown in Fig.~\ref{fig:m:t:m2:md2}(b).
We find that $\overline{\langle m^2\rangle}$ scales as $1/N$ which confirms that fluctuations are a finite-size effect (we actually find numerically $\overline{\langle m^2\rangle} = 1/N$ to a good accuracy).
However, the presence of temporal oscillations can be revealed by considering a specific, disorder-dependent observable.
We introduce for each site $i$ new variables $w_i$ and $v_i$ defined as
\be\label{eq:change:var} w_i=\epsilon_i s_i \quad \text{and} \quad v_i=\epsilon_ih_i,\ee
 and the corresponding macroscopic observables,
\be
w = \frac{1}{N}\sum_{i=1}^N w_i \quad \text{and} \quad v = \frac{1}{N}\sum_{i=1}^N v_i.
\ee 
Once expressed in terms of the variables $w_i$ and $v_i$, the interaction energies $E_s$ and $E_h$ defined in Eqs.~(\ref{eq:energy:Es}) and (\ref{eq:energy:Eh}) take the same form as the ones of the model without disorder (i.e., with $\epsilon_i=1$ for all $i$), upon replacement of the observables $(m, h)$ by $(w, v)$.
Hence, in these new variables, the model is the same as the model without disorder. For $T<T_c$, $w(t)$ oscillates with time with a nonzero amplitude, as depicted in Fig.~\ref{fig:m:t:m2:md2}.
As oscillations are not visible on the magnetisation, but can be revealed by using a non-trivial, disorder-dependent observable, we call such oscillations `hidden oscillations'. 
In addition, the change of variable can be used to justify why the magnetisation no longer carries a signature of temporal oscillations. 
In practice, the disorder introduced here consists of separating the $N$ sites into two subsets: a subset $S_+$ where $\epsilon_i=+1$ and a subset $S_-$ where $\epsilon_i=-1$. 
Both subsets have an average size of $N/2$. We introduce a local magnetisation for each subset as
\be w_{\pm}=\frac{2}{N}\sum_{i\in S_{\pm}}w_i,\ee which results in
\be\label{eq:decomposition:m:wplus:wmoins} m(t)=\frac{1}{2}\left[ w_{+}(t)-w_{-}(t)\right].\ee
As the interactions are identical for all $w_i$ and all $v_i$, we have in the limit of a large system size $N$, $w^+\sim w^-\sim w$, so that $\langle m^2\rangle \to 0$.
According to this heuristic argument, the fluctuations observed on Fig.~\ref{fig:m:t:m2:md2} are thus expected to be finite-size effects.

\begin{figure}
    \centering
    \includegraphics{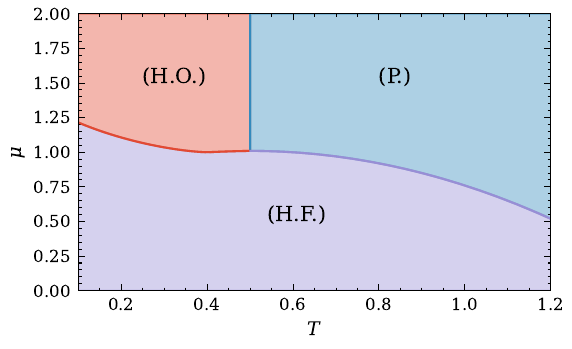}
    \caption{Phase diagram of the model for $J_1=0.6$ and $J_2=0.4$. Three distinct behaviours are observed: a paramagnetic phase (P.), a hidden ferromagnetic phase (H.F.) and a phase with hidden oscillations (H.O.).}
    \label{fig:phase_diagram}
\end{figure}
The phase diagram of the model, displayed in Fig.~\ref{fig:phase_diagram}, can be obtained from the phase diagram of the model without disorder \cite{guislain_nonequil2023,guislain_discontinuous2024}
using the change of variable given in Eq.~(\ref{eq:change:var}). 
The model exhibits three distinct phases: a paramagnetic phase at high temperature $T$, a `hidden ferromagnetic' phase at low $T$ and low $\mu$, and a hidden oscillating phase at
low $T$ and high $\mu$. By `hidden ferromagnetic' phase, we mean that no spontaneous magnetisation is visible, but the disorder-dependent observable $w$ takes a non-zero static value.
This phase exhibits some similarities with a spin-glass phase, but at odds with a genuine spin-glass phase, the overlap distribution remains that of a ferromagnet (see Sec.~\ref{sec:overlap} for the definition of the overlap).


\section{Linear and non-linear generalised susceptibilities}

\subsection{Linear susceptibility}

\begin{figure}
    \centering
    \includegraphics{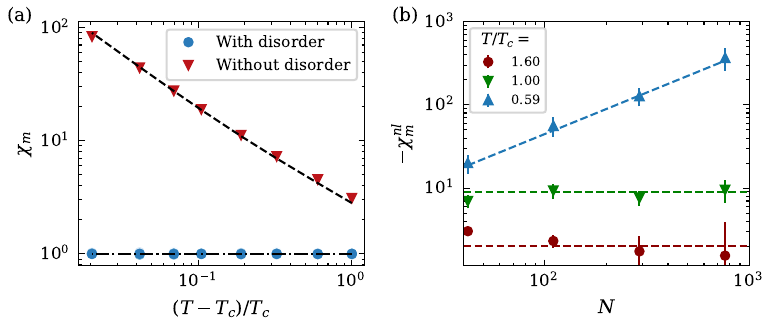}
    \caption{(a) Linear susceptibility $\chi_m$ versus $(T-T_c)/T_c$ for the model without disorder (for $N=10^5$, red triangles), and the model with disorder (blue circles) for systems size from $N=16$ to $761$. The difference between various system sizes is small and data points cannot be distinguished. The dashed lines correspond to the analytical predictions Eq.~(\ref{eq:chi:l:wd}) and Eq.~(\ref{eq:chi:l:d}). 
    (b) Non-linear susceptibility (with a minus sign) $-\chi_{m}^{nl}$ versus $N$ for different temperatures: red circles $T=1.6T_c$, 
    green triangles $T=T_c$, blue triangles $T=0.59T_c$. 
    The non-linear susceptibility $\chi_{m}^{nl}$ diverges only for $T>T_c$. Parameters: $J_1=0.6$, $J_2=0.4$, $\mu=1.4$.}
    \label{fig:chi:linear:non:linear:m}
\end{figure}
Beside the onset of a non-zero value of the order parameter, a phase transition can also be characterized by the divergence of a susceptibility associated with the order parameter when approaching the transition from the symmetric phase (i.e., the phase where the order parameter is zero, like the paramagnetic phase).
In an Ising model at equilibrium with $N$ spins and an external field $h$, the linear susceptibility $\chi=dm/dh|_{h=0}$ can be linked to fluctuations of the magnetisation
through the static fluctuation-dissipation relation, $\chi= N \langle m^2\rangle$. This suggests to introduce in the non-equilibrium disordered model a generalised susceptibility $\chi_{m}$ as \be \label{eq:chi:l:def}\chi_m= N\overline{ \langle m^2\rangle}. \ee
In the model without disorder for $T>T_c$, one finds for $N\gg1$
\be \label{eq:chi:l:wd}
\chi_{m}^{0}=\frac{2}{av_0\sqrt{\pi}}\left(\frac{T-T_c}{T_c}\right)^{-1}, \qquad T \to T_c^{+}
\ee
($v_0$, $a$ are given in \ref{appendix:coeff:lettre}) such that the susceptibility $\chi_m^0$ diverges as $(T-T_c)^{-1}$ when approaching the critical point from the paramagnetic phase
[see Fig.~\ref{fig:chi:linear:non:linear:m}(a)].
In the model with disorder, using the change of variable given in Eq.~(\ref{eq:change:var}), we have
\be \overline{\langle m^2\rangle}=\frac{1}{N^2}\sum_{i, j}\overline{\epsilon_i\epsilon_j}\langle w_i w_j\rangle. \ee 
The variables $\epsilon_i$ are uncorrelated random variables satisfying $\overline{\epsilon_i\epsilon_j}=\delta_{i,j}$, so that 
\be\label{eq:m2:av} \overline{\langle m^2\rangle}=N^{-1}\ee 
as observed in Fig.~\ref{fig:m:t:m2:md2} (we have used that $w_i^2=1$). From Eq.~(\ref{eq:chi:l:def}), one finds
\be \label{eq:chi:l:d}\chi_{m}=1\ee
for temperatures below and above $T_c$, as reported in Fig.~\ref{fig:chi:linear:non:linear:m}(a). The linear susceptibility no longer diverges at $T_c$, similarly to the behaviour
of the original Mattis model \cite{mattis1976}. However, a divergence can be observed by considering non-linear susceptibilities.

\subsection{Non-linear susceptibility}

In a spin-glass context, the presence of a hidden phase transition may be revealed by considering the non-linear susceptibility $\chi_{nl}=\overline{d^3m/dh^3}|_{h=0}$ \cite{mattis1976}.
At equilibrium, the non-linear susceptibility $\chi_{nl}$ is linked to the fourth-order cumulant of magnetisation fluctuations, $\chi_{nl}=\beta^3N^3\left(\langle m^4\rangle -3\langle m^2\rangle^2\right)$.
We thus introduce in the non-equilibrium disordered model a generalised non-linear susceptibility associated with the magnetisation, as
\be \chi_{m}^{nl}=N^3\overline{\left(\langle m^4\rangle -3\langle m^2\rangle^2\right)}.\ee  
%
Using the change of variables introduced in Eq.~(\ref{eq:change:var}) as well as the relation 
\be \overline{\epsilon_i\epsilon_j\epsilon_k\epsilon_l}=\delta_{i, j}\delta_{k, l}+\delta_{i, k}\delta_{j, l}+\delta_{i, l}\delta_{j, k}-2\delta_{i,j,k,l},\ee
one finds (see \ref{appendix:chi:nl:m:calcul})
\be \chi_{m}^{nl}=-2-\frac{6}{N}\sum_{i\neq j}\langle w_iw_j\rangle^2.\ee
Using that all $\langle w_iw_j\rangle$ are equal for $i\neq j$,   
one gets (see \ref{appendix:chi:nl:m:calcul}),
\be \chi_{m}^{nl}= -2-6N\left(\langle w^2\rangle-N^{-1}\right)^2.\ee
In the limit $N\to\infty$, one finds that for $T>T_c$, $\chi_{m}^{nl}=-2$ whereas at $T=T_c$,
\be \label{eq:chim:nl:crit}
\chi_{m}^{nl} = -2-6\gamma^2,
\ee
where $\gamma=(v_0 \sqrt{\pi b})^{-1}$ (see \ref{appendix:coeff:lettre} for the values of $b$ and $v_0$). 
The non-linear susceptibility $\chi_{m}^{nl}$ is obtained numerically and plotted in Fig.~\ref{fig:chi:linear:non:linear:m}(b) for different temperatures around $T_c$.
It does not diverge at $T=T_c$, but is discontinuous at $T_c$ in the $N\to\infty$ limit. Moreover, it diverges with $N$ below the transition ($T<T_c$).
For finite-size systems, the discontinuity is smoothed. 
This is different from the behaviour observed in spin glasses, where the non-linear susceptibility has a jump or diverges at $T_c$ but remains finite on both sides of the transition \cite{fujiki.katsura.1981,suzuki.miyashita.1981}. 
We obtain a behaviour similar to that observed in the equilibrium Mattis model \cite{mattis1976,aharony.imry.1976}: the non-linear susceptibility diverges when approaching $T_c$ from the paramagnetic phase in dimension three or lower, but it does not diverge in dimension four and above, and thus not in mean field. However, for $T<T_c$, the non-linear susceptibility diverges with $N$ in all dimensions.

To sum up, we have shown that the magnetisation no longer carries a signature of a phase transition when separable disorder is added to the model. Yet, because of the simple structure of the disorder, a change of spin variables enables oscillation detection.
In the absence of disorder, the magnetisation is the natural order parameter associated with spin reversal symmetry breaking. However, to distinguish between a paramagnetic to a hidden ferromagnetic phase transition and a paramagnetic to a hidden oscillating phase transition, the statistical characterisations obtained from this order parameter are not sufficient to describe the phase transition.
In the following section, we thus characterise the transition to oscillations using other observables that specifically reveal the non-equilibrium character of the oscillating state.
At odds with the magnetisation, though, such non-equilibrium observables require the explicit knowledge of the variables $\epsilon_i$ defining the disorder.


\section{Non-equilibrium characterizations of the hidden phase transition}

\subsection{Stochastic time derivative of the magnetisation}
\label{sec:time:deriv:mag}

When the magnetisation oscillates in time, time-translation invariance is broken. The order parameter associated with this symmetry is the stochastic derivative of the magnetisation defined in \cite{guislain_nonequil2023} as 
\be \label{eq:def:mdot} \dot{m}(\mC)=-\frac{2}{N}\sum_{i=1}^Ns_i W_1^{s_i}(\mC), \ee
where $W_1^{s_i}$ is the transition rate corresponding to a flip of spin $s_i$ defined in Eq.~(\ref{eq:transition:rates}) and
$\mC=\{s_i, h_i\}_{i=1,\dots,N}$ is a given configuration of the spins and fields. 
Unlike the magnetisation, this quantity cannot be directly determined from the spin configuration, as it requires the knowledge of the disorder.
However, in numerical simulations, the disorder is known explicitly, and the stochastic time derivative of the magnetisation can thus be determined.
We plot in Fig.~\ref{fig:m:t:m2:md2} $\overline{\langle \dot{m}^2\rangle}$ for a temperature $T<T_c$, and observe that it decreases with system size.
Hence the quantity $\overline{\langle \dot{m}^2\rangle}$ exhibits no signature of the broken time-translation invariance for $T<T_c$ in the disordered model
(which is not surprising given that the magnetisation does not oscillate).

In the previous section, we showed, using the change of variables introduced in Eq.~(\ref{eq:change:var}), that the energies $E_1$ and $E_2$ [Eqs.~(\ref{eq:energy:Es}) and (\ref{eq:energy:Eh})]
do not explicitly depend on the disorder once expressed as a function of $w, v$. We thus introduce 
\be G_1^{w_i}(w, v)=\left[1+e^{2\beta w_i\left(J_1w+ v\right)}\right]^{-1},\ee
\be G_2^{w_i}(w, v)=\left[1+e^{2\beta v_i\left(J_2v+ (1-\mu)w\right)}\right]^{-1}.\ee
We find:
\be \dot{m}=-\frac{2}{N}\sum_{i=1}^N \epsilon_i w_i G_1^{w_i}(w, v).\ee
Averaging $\dot{m}^2$ over microscopic configurations and over the disorder, we find (see \ref{appendix:mdot2:calcul} for details)
\begin{equation}\label{eq:md2:av}\overline{\langle \dot{m}^2\rangle}=N^{-1}\left(1-\langle w^2\rangle +\langle \dot{w}^2\rangle\right)
\end{equation}
where $\dot{w}=-2N^{-1}\sum_{i=1}^Nw_i G_1^{w_i}(w,v)$ corresponds to the stochastic derivative of the magnetisation of the model without disorder,
upon replacement of the observables $(m,h)$ by $(w,v)$.
At leading order in $N$, one finds that $\overline{\langle \dot{m}^2\rangle}=N^{-1}$ for $T\geq T_c$. For $T<T_c$, the terms $\langle w^2\rangle$ and $\langle \dot{w}^2\rangle$ have a finite nonzero limit with $N\to\infty$. 
Thus, for all temperatures, $\overline{\langle \dot{m}^2\rangle}$ decreases as $N^{-1}$ and similarly to the magnetisation $m$,
the stochastic derivative of the magnetisation $\dot{m}$ is no longer an order parameter when separable disorder is included into the model. 

\begin{figure}[t]
    \centering
    \includegraphics{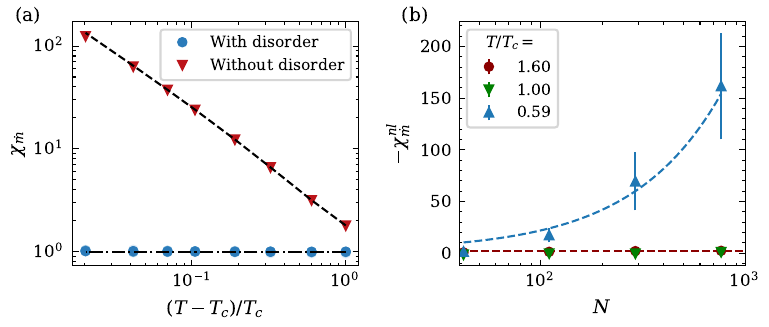}
    \caption{(a) Linear susceptibility $\chi_{\dot{m}}$ as a function of $(T-T_c)/T_c$ for the model without disorder ($\chi_{\dot{m}}^0$, red triangles for $N=10^5$) and the model with disorder (blue circles, for different system sizes $N\in[50,100, 500, 1000, 2000]$; the difference between various system sizes is small and the different points cannot be distinguished). Lines correspond to the analytical predictions, Eq.~(\ref{eq:chi:l:md:wd}) for the model without disorder (dashed line) and Eq.~(\ref{eq:chi:l:md:d}) for the model with disorder (dot-dashed line), which here boils down to $\chi_{\dot{m}}=1$. 
    (b) Non-linear susceptibility (with a minus sign) $-\chi_{\dot{m}}^{nl}$ as a function of $N$ for different temperatures (see legend).
    The non-linear susceptibility $\chi_{\dot{m}}^{nl}$ diverges only for $T>T_c$.
    Parameters: $J_1=0.6$, $J_2=0.4$, $\mu=1.4$.}
    \label{fig:chi:linear:non:linear:md}
\end{figure}

One can also define a generalised linear susceptibility associated with the order parameter $\langle \dot{m}^2\rangle$ as, \be \chi_{\dot{m}}=N\langle\dot{m}^2\rangle.\ee
In the model without disorder, one finds that 
\be \label{eq:chi:l:md:wd}
\chi_{\dot{m}}^{0} = \frac{2}{a\sqrt{\pi}}\left(\frac{T-T_c}{T_c}\right)^{-1}, \qquad T \to T_c^{+},
\ee
so that $\chi_{\dot{m}}^{0}$ diverges as $(T-T_c)^{-1}$ [see Fig.~\ref{fig:chi:linear:non:linear:md}(a)].
Using Eq.~(\ref{eq:md2:av}) and the fact that close to the transition $\langle w^2\rangle \ll 1$ and $\langle \dot{w}^2\rangle\ll 1$, one instead finds in the model with separable disorder that
\be \chi_{\dot{m}}=1\ee
for $T\lesssim T_c$. The susceptibility $\chi_{\dot{m}}$ in the presence of disorder no longer diverges at $T_c$, as depicted in Fig.~\ref{fig:chi:linear:non:linear:md}(a).
%

Moreover, like for the magnetisation, one can also define a non-linear susceptibility associated with $\dot{m}$ as, 
\be\label{eq:chi:nl:mdot:def} \chi_{\dot{m}}^{nl}=N^3\overline{\left(\langle \dot{m}^4\rangle -3\langle \dot{m}^2\rangle^2\right)}.\ee
Keeping only terms that contribute in the large-$N$ limit, one finds (see \ref{appendix:chi:nl:md:calcul})
\be\label{eq:chi:l:md:d} \chi_{\dot{m}}^{nl}=3N\left(\langle \dot{w}^4\rangle -3\langle \dot{w}^2\rangle^2+\langle w^4\rangle -\langle w^2\rangle^2 \right) -2.\ee
In the limit $N\to\infty$, one finds $\chi_{\dot{m}}^{nl}=-2$ for $T>T_c$ and $\chi^{nl}_{\dot{m}}=-2+\rho$ at $T=T_c$, where $\rho$ is given by:
\be \label{eq:rho:value}
\rho=3\left(\frac{-4+\pi}{\pi b }+\frac{-2+\pi}{v_0^2\pi b }\right).
\ee
[see \ref{appendix:chi:nl:md:calcul} for the derivation of Eq.~(\ref{eq:rho:value}), and \ref{appendix:coeff:lettre} for the values of $b$ and $v_0$].
Like for the magnetisation, the non-linear susceptibility associated with the stochastic derivative of the magnetisation diverges with $N$ below $T_c$,
indicating the presence of a phase transition. However, it does not diverge when approaching $T_c$ from above, because of the mean-field nature of the model.
The non-linear susceptibility is discontinuous at $T_c$ in the limit $N\to\infty$.

\subsection{Entropy production}

In the framework of stochastic thermodynamics \cite{seifert_stochastic_2012}, the transition to a limit cycle may also be characterized as a transition from microscopic to macroscopic irreversibility, by introducing the entropy production density $\sigma=\Sigma/N$ in the limit $N\to\infty$ \cite{xiao_entropy_2008,seara_irreversibility_2021}. The steady-state entropy production $\Sigma$ identifies with the entropy flux \cite{gaspard_time-reversed_2004}:
\be
\Sigma = \sum_{\mC\neq \mC'}  W(\mC'|\mC)P(\mC)\, \ln \frac{W(\mC'|\mC)}{W(\mC|\mC')} \,,
\ee
where $\mC=\{s_i, h_i\}_{i=1,\dots,N}$ is a microscopic configuration of the spins and fields.
As there are two types of transitions ($k=1$ corresponding to a spin flip, and $k=2$ to a field reversal), we have
\be
\Sigma = \sum_{\mC=\{s_i, h_i\}} \sum_{i=1}^N \left[W_1^{s_i}(\mC) \, P(\mC)\, \ln \frac{W_1^{s_i}(\mC)}{W_1^{-s_i}(\mC^{s_i})} +W_2^{h_i}(\mC)\, P(\mC)\, \ln \frac{W_2^{h_i}(\mC)}{W_2^{-h_i}(\mC^{h_i})}\right]\,,
\ee
where $\mC^{s_i}$ and $\mC^{h_i}$ are obtained from the configuration $\mC=\{s_i, h_i\}_{i=1,\dots,N}$ after flipping the spin $s_i$ or the field $h_i$.
For the model with separable disorder, applying the change of variable given in Eq.~(\ref{eq:change:var}) leads to the same expression of the entropy production $\Sigma$
as for the model without disorder. We thus have 
\be
\Sigma \!=\!\!\! \sum_{\mC=\{w_i, v_i\}} \sum_{i=1}^N P(\mC)\left[G_1^{w_i}(w, v) \ln \frac{G_1^{w_i}(w, v)}{G_1^{-w_i}(w-\frac{2w_i}{N}, v)}+G_2^{v_i}(w, v) \ln \frac{G_2^{v_i}(w, v)}{G_2^{-v_i}(w, v-\frac{2v_i}{N})}\right] \,
\ee
where we recall that $w=N^{-1}\sum_i w_i$ and $v=N^{-1}\sum_{i=1}^N v_i$.
We then change the sum over configurations into integrals over $w$ and $v$ together with a restricted sum: $\sum_{\mC}=\int dw dv\sum_{\mC\in S(w, v)}$, where $S(w, v)$ denotes the set of configurations
$\mC=\{w_i, v_i\}_{i=1,\dots,N}$ with $w(\mC)=w$ and $v(\mC)=v$. We note $Nn_k^{\sigma}$ the number of possible transitions $(k, \sigma)$ ($k=1, 2$ for a spin or a field, and $\sigma$ corresponds to the sign of the corresponding spin or field variable), from a configuration $\mC=\{w_i, v_i\}_{i=1,\dots,N}$,
\be
n_1^{\pm} = \frac{1}{2}(1\pm w), \quad  n_2^{\pm} = \frac{1}{2}(1\pm v).
\ee
We note $\sigma\mathbf{d}_k$ the jumps $(\Delta w, \Delta v)$ for a transition of type $(k,\sigma)$. We have $\mathbf{d}_1=(-2/N, 0)$ and $\mathbf{d}_2=(0, -2/N)$.
We find for the entropy production
\be \Sigma=N\sum_{k\in\{1,2\}}\sum_{\sigma=\pm1}\left\langle  n_k^{\sigma} G_k^{\sigma}(w, v)\ln\left[\!\frac{G_k^{\sigma}(w, v)}{G_k^{-\sigma}((w,v)+\frac{\sigma \mathbf{d}_k}{N})} \right]\!\right\rangle,\ee
with the notation $\langle A(w, v)\rangle=\int dwdv\, P_N(w, v)A(w, v)$. 
To leading order in $N$, we find,  
\be \Sigma=N\sum_{k\in\{1,2\}}\left\langle \left(n_k^+G_k^{+}-n_k^-G_k^{-}\right)\ln \left[\frac{G_k^{+}n_k^{-}}{G_k^{-}n_k^{+}}\right]\right\rangle,\ee
where we make the dependence in $(w,v)$ implicit. We define
\be
\Sigma_k=\left\langle \left(n_k^+G_k^{+}-n_k^-G_k^{-}\right)\ln \left[G_k^{+}/G_k^{-}\right]\right\rangle.
\ee
For $k=1$, we have $n_1^+G_1^+-n_1^-G_1^-=\dot{w}/2$ with
\be
\dot{w}=-w+\tanh[\beta(J_1w+v)]
\ee
and
\be
\ln \left[G_1^{+}/G_1^{-}\right]=2\tanh^{-1}(w+\dot{w}).
\ee
We thus find
\be
\Sigma_1=N\langle \dot{w}\tanh^{-1}(w+\dot{w})\rangle.
\ee
\begin{figure}
    \centering
    \includegraphics{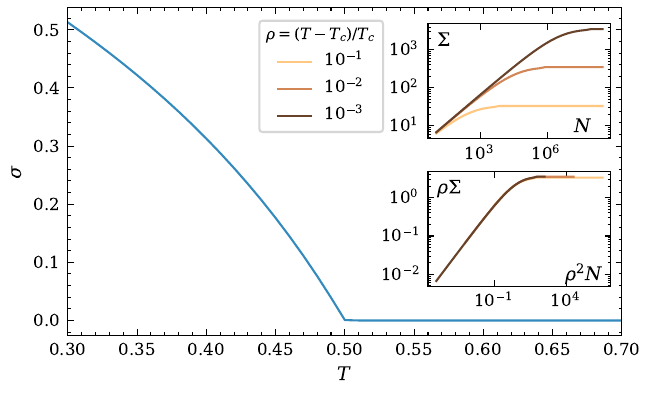}
    \caption{Entropy production density $\sigma$ in the limit $N\to\infty$ versus temperature $T$. Insets: entropy production $\Sigma$ versus system size $N$ (top inset) and rescaled entropy production $\rho \Sigma$ versus rescaled system size $\rho^2N$ (bottom inset) for different reduced temperatures $\rho=(T-T_c)/T_c$.
    Parameters: $J_1=0.6$, $J_2=0.4$, $\mu=1.4$.}
    \label{fig:entropy:prod}
\end{figure}
Close to the transition, $w$ and $\dot{w}$ are small, so that $\tanh^{-1}(w+\dot{w})\sim w+\dot{w}$. 
In addition, we know that $\langle w\dot{w}\rangle=0$ \cite{guislain_nonequil2023}. We thus obtain
\be\label{eq:sigma1} \Sigma_1/N=\langle \dot{w}^2\rangle.\ee
For $k=2$, we introduce
\be
\dot{v}=-v+\tanh[\beta(J_2v+(1-\mu)w)],
\ee
the equivalent of $\dot{w}$ for the field variables $v_i$. Then, we find $\Sigma_2=N\langle \dot{v}\tanh^{-1}(v+\dot{v})\rangle$. As spins and fields play a symmetric role in this model, one also finds \be\label{eq:sigma2} \Sigma_2/N=\langle \dot{v}^2\rangle.\ee
In the paramagnetic phase of the model without disorder ($T>T_c$), one has $\langle\dot{w}^2\rangle \sim \langle\dot{v}^2\rangle \sim (T-T_c)^{-1}N^{-1}$, resulting in $\Sigma\sim (T-T_c)^{-1}$. In contrast, for $T<T_c$, one has $\langle \dot{w}^2\rangle\sim \langle \dot{v}^2\rangle\sim (T_c-T)$ so that $\Sigma\sim N(T_c-T)$.
The entropy production diverges at $T_c$ for $T>T_c$ (see insets of Fig.~\ref{fig:entropy:prod}, with and without rescaling) and becomes proportional to the system size for $T<T_c$.
The entropy production density $\sigma=\Sigma/N$ is plotted in the main panel of Fig.~\ref{fig:entropy:prod}, in the limit $N \to \infty$, showing that $\sigma>0$ for $T<T_c$ and
$\sigma=0$ for $T>T_c$.
Hence, the entropy production density can be considered as an order parameter revealing the phase transition to an oscillating phase, both in the presence and in the absence of disorder.


\section{Overlap distribution}
\label{sec:overlap}

\begin{figure}[t]
    \centering
    \includegraphics{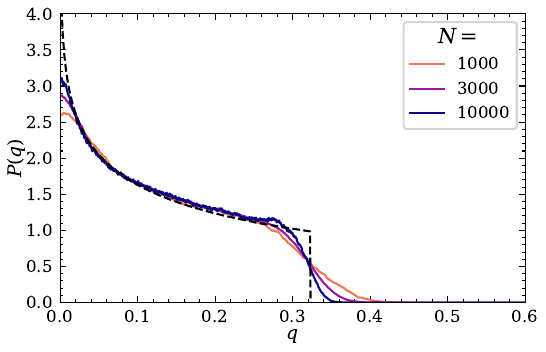}
    \caption{Overlap distribution $P(q)$ for different system sizes $N$, showing the convergence to the infinite-$N$ limit (black dashed line). 
    Parameters: $T=0.8T_c$, $\mu=1.4$, $J_1=0.6$, $J_2=0.4$.  
    }
    \label{fig:overlap}
\end{figure}

We have shown that the model with Mattis-type separable disorder presents a hidden phase transition, where the oscillations can be revealed using a simple change of variable which requires the knowledge of the disorder. However, one may wonder how to detect the presence of oscillations if one only has access to spin configurations, and not to the configuration of disorder.
In particular, we have seen that the magnetisation carries no signature of the transition to temporal oscillations.
In this section, we show that the overlap distribution, which can be determined from the sole knowledge of spin configuration statistics, may be used as an order parameter of the transition
to spontaneous oscillations.
We thus consider the overlap
\be
q_{ab} = \frac{1}{N} \sum_{i=1}^N s_i^{a} s_i^{b}
\ee
between two statistically independent spin configurations $\{s_i^{a}\}$ and $\{s_i^{b}\}$.
Overlaps have been introduced in spin-glass models to deal with the absence of a visible order in the spin-glass phase \cite{mezard_spin_1987}. Identical (opposite) configurations have an overlap $q_{ab}=1$ ($q_{ab}=-1$), while $q_{ab}=0$ for uncorrelated configurations.
We consider the overlap distribution
\be P(q)=\sum_{\{s_i^a\}, \{s_i^b\}}\overline{P(\{s_i^a\})P(\{s_i^b\})\, \delta\left(\frac{1}{N}\sum_{i=1}^Ns_i^as_i^b-q\right)}.\ee
The overlap distribution $P(q)$ is plotted in Fig.~\ref{fig:overlap} for $T<T_c$, showing convergence to the infinite-size distribution when $N$ increases.
We observe a non-trivial structure of the overlap distribution as it is spread over a continuous interval. This property is usually considered as a hallmark of a continuous replica symmetry breaking in the spin-glass context \cite{mezard_spin_1987}.
However, this non-trivial structure is not due here to a replica symmetry breaking but to the breaking of the time-translation invariance in the oscillating phase. 
Indeed, as the change of variable from $s_i$ to $w_i$ 
amounts to a permutation of configurations, one has $P(\{s_i\})=P(\{w_i\})$ and $\sum_{i=1}^Ns_i^as_i^b =\sum_{i=1}^Nw_i^aw_i^b$.
Hence one obtains 
\be
P(q)=\sum_{\{w_i^a\}, \{w_i^b\}}P(\{w_i^a\})P(\{w_i^b\})\, \delta\left(\frac{1}{N}\sum_{i=1}^Nw_i^aw_i^b-q\right).
\ee
This expression is identical to the overlap distribution for the model without disorder defined in terms of spins $w_i$. 
The overlap distribution can be obtained analytically for small $T-T_c$ in the limit $N \to \infty$ (see \cite{guislain_nonequil2023,guislain2024overlap}).
For $T>T_c$ the overlap distribution is a delta-peak in $q=0$ whereas for $T<T_c$, it is non-zero over a continuous interval $[-q_0, q_0]$.

\section{Discussion}

In this section, we briefly discuss two relatively straightforward generalizations of the disordered mean-field spin model considered above.
We first discuss in Sec.~\ref{sec:partial:knowledge} the issue of oscillation detection when the disorder is only partially known.
Then, we generalize the model in Sec.~\ref{app:interpol:disorder} to continuously interpolate between the non-disordered model and the disordered model with separable disorder.

\subsection{Partial knowledge of disorder}
\label{sec:partial:knowledge}

\begin{figure}[t]
    \centering
    \includegraphics{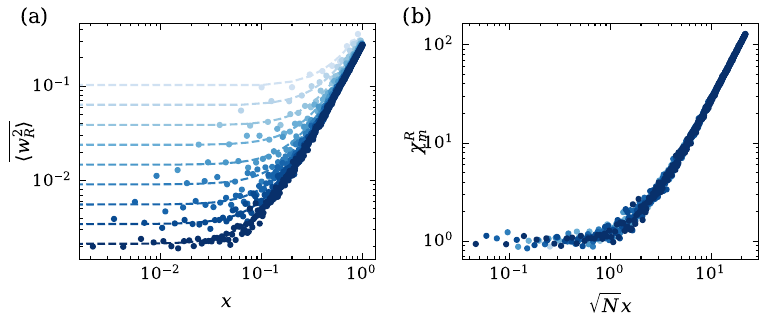}
    \caption{(a) $\langle w_R^2\rangle$ versus $x$ for different $N$ and (b) $\chi_{m}^R=N\overline{\langle w_R^2\rangle}$ versus $\sqrt{N}x$ for $N\in[10, 16, 26, 43, 68, 110, 180, 290, 470]$ (from lighter to darker colours).
    Parameters: $T=0.59T_c$, $\mu=1.4$, $J_1=0.6$, $J_2=0.4$. 
    }
    \label{fig:partial:disorder}
\end{figure}

Assuming that the variables $\epsilon_i$ characterising the disorder are known explicitly, we can construct an observable $w(t)$ which contains all information on the time-oscillating structure,
as discussed in Sec.~\ref{sec:hidden:oscill}.
However, one may wonder how to detect oscillations in the case when the disorder variables $\epsilon_i$ are not known explicitly, similarly to what happens in experiments where disordered coupling constants are not accessible. We have seen in Sec.~\ref{sec:overlap} that oscillations can be detected in an indirect way by measuring the overlap distribution, which requires only the knowledge of spin configurations, but not of the disorder. Here, we address a slightly different question: imagine that the disorder variables $\epsilon_i$ are not known explicitly, but can be reconstructed through some noisy measurement process.
In this case, one does not have access to the true variables $\epsilon_i$, but only to the reconstructed variables $\epsilon^R_i$ obtained from the measurement procedure, which only partially correspond to $\epsilon_i$.
We define the reconstructed observable $w_R$ as
\be w_R = \frac{1}{N} \sum_{i=1}^N\epsilon^R_i s_i.\ee
Averaging over spin configurations and over disorder, one finds
\be \overline{\langle w_R^2\rangle}=\langle w^2\rangle \text{cor}(\epsilon, \epsilon^R)^2+N^{-1}(1-\text{cor}(\epsilon, \epsilon^R)^2)\ee
where $\text{cor}(\epsilon, \epsilon^R)=N^{-1}\sum_{i=1}^N\overline{\epsilon_i\epsilon^R_i}$ is the correlation between the exact $\epsilon=\{\epsilon_i\}$ and the measured
$\epsilon^R=\{\epsilon_i^R\}$.
We assume that each variable $\epsilon_i^R$ obtained from the measurement process is such that $\epsilon_i^R=\epsilon_i$ with a probability $(1+x)/2$ and $\epsilon_i^R=-\epsilon_i$ with a probability $(1-x)/2$. The case $x=1$ corresponds to a noiseless measurement, while for $x=0$ the output of the measurement is completely random.
With this notation, it follows that $\text{cor}(\epsilon, \epsilon^R)=x$, and one obtains
\be\label{eq:res:wr2}
\overline{\langle w_R^2\rangle}=x^2\langle w^2\rangle  +N^{-1}(1-x^2).
\ee
This quantity $\overline{\langle w_R^2\rangle}$, obtained from numerical simulations, is plotted in Fig.~\ref{fig:partial:disorder}(a) for different system sizes, at a temperature $T<T_c$.
In addition, we introduce the reconstructed susceptibility 
\be \chi_{m}^R=N\overline{\langle w_R^2\rangle},\ee 
and we find
\be  \chi_m^R=Nx^2\langle w^2\rangle +1.\ee
For $T<T_c$, $\langle w^2\rangle$ is independent of $N$, thus $\chi_m^R$ plotted as a function of $\sqrt{N}x$ is independent of $N$, as seen in Fig.~\ref{fig:partial:disorder}(b).
For very low values of $x$, one finds $\chi_m^R\sim 1$ which is independent of whether or not hidden oscillations are present as 
$\chi_m^R$ takes the same value as for $T>T_c$. For higher values of $x$, one finds $\chi_m^R \sim (\sqrt{N}x)^2\langle w^2\rangle$ which contains a signature of the hidden oscillations through the value of $\langle w^2\rangle$. The transition between the two regimes takes place when $(\sqrt{N}x)^2\langle w^2\rangle \sim 1$.
Since $\langle w^2\rangle\sim (T_c-T)$, oscillations can be detected when 
\be x\gtrsim  \frac{1}{\sqrt{N}(T_c-T)}.\ee
Hence, for a large system and not too close to the critical temperature, evidence for oscillations can be obtained even
if the disorder is determined through a very noisy measurement process ($x \ll 1$).

\subsection{Biased disorder}
\label{app:interpol:disorder}

We finally turn to a generalization of the model aimed at interpolating between the non-disordered and disordered models.
We consider the same type of disorder as before but we now assume that $\epsilon_i$ follows a Bernouilli distribution: $\epsilon_i=+1$ with a probability $p$ and $\epsilon_i=-1$ with a probability $1-p$. The previous model with separable disorder corresponds to $p=1/2$. The cases $p=1$ and $p=0$ correspond to the model without disorder.

As done in Eq.~(\ref{eq:decomposition:m:wplus:wmoins}), we split the magnetisation $m$ into two contributions, one where $\epsilon_i=+1$ and one where $\epsilon_i=-1$
\be m=pw_+-(1-p)w_-,\ee
with 
\be w_{+}=\frac{1}{pN}\sum_{i|\epsilon_i=+1}w_i, \quad \quad w_{-}=\frac{1}{(1-p)N}\sum_{i|\epsilon_i=-1}w_i.\ee
In the limit of large system size, $w_+\sim w_-\sim w$, we thus obtain that
\be m(t)\sim (2p-1)w(t).\ee
For $p\neq 1/2$, we observe a transition to temporal oscillations of the magnetisation $m(t)$ still at $T=T_c$, but the amplitude of the oscillations depends on $p$ as 
\be \overline{\langle m^2\rangle}=(2p-1)^2 \langle w^2\rangle.\ee
We recover that for $p=1/2$, the magnetisation no longer exhibits temporal oscillations.
For all values of $p$, though, the full oscillations can be recovered in terms of the variable $w=N^{-1}\sum_i w_i$, with $w_i=\epsilon_is_i$.

\section{Conclusion}

To conclude, we have studied a disordered spin model with non-reciprocal interactions, which exhibits hidden oscillations that are only visible using a specifically tailored
disorder-dependent observable.
The model is built by incorporating separable (Mattis-type) quenched disorder in a model known to exhibit spontaneous temporal oscillations at low temperature and with strongly
non-reciprocal interactions \cite{guislain_nonequil2023,guislain_discontinuous2024}. Due to the simple structure of the disorder, a mapping to the non-disordered model can be used to evaluate
different observables analytically.
While magnetisation or linear susceptibilities provide no indication of a phase transition, non-linear (third-order) generalised susceptibilities yield clear signatures of the onset of oscillations.
We have also shown that the presence of oscillations implies a non-zero entropy production density, as well as a non-trivial overlap distribution with a continuous support, somewhat reminiscent of a continuous replica symmetry breaking, but with a different physical origin.

A specificity of separable  disorder is the absence of frustration.
As for future work, it would be of interest to explore whether hidden oscillations could also exist, and be detected, in the present spin model with more general forms of disordered couplings
which generate frustration. More generally, the competition between frustration and non-reciprocity of interactions in far-from-equilibrium heterogeneous systems is a topic that certainly deserves further exploration in a broader context, like in neural network dynamics \cite{pina2018,kalitzin2014} or in population dynamics with many interacting species \cite{ros2023}.



\appendix
\section{Coefficients}
\label{appendix:coeff:lettre}

We provide here the values of the parameters $v_0$, $a$ and $b$ introduced in Eqs.~(\ref{eq:chi:l:wd}) and (\ref{eq:rho:value}):
\begin{align} &v_0=[\mu-\mu_0(T)]/T^2,\\
&a=\frac{ 2 TT_c}{T^2D_{22}+\mu-\mu_0(T)},\\
&b=\frac{a_1 T^4+3T^2[\mu-\mu_0(T)]a_3}{4[\mu-\mu_0(T)][T^2D_{22}+\mu-\mu_0(T)]},\end{align}
where the coefficients appearing in the above expressions are given by
\begin{align}&\mu_0(T)=1-T^2(1-J_1/T)(1-J_2/T),\\
&a_{1}=\! (2 J_2 + J_2^3 + 3 J_1)/T -2 \!+\!\! J_2 (-1 + J_1 J_2 +\mu)^2/T^3\!-\!\! 2(1 + J_2^2) (-1 + J_1 J_2 +\mu)/T^2,\\
 &a_{3}=-2/3 +(2J_2 + J_2^3 + 3J_1)/3T,\\
& D_{22}=1/T^2+(J_1/T-1)^2.\end{align}


\section{Computation of $\overline{\langle \dot{m}\rangle}$}
\label{appendix:mdot2:calcul}
From the definition of $\dot{m}$ given in Eq.~(\ref{eq:def:mdot}), we have 
\be \overline{\langle \dot{m}^2\rangle}=\left< \frac{4}{N^2}\sum_{i, j}\overline{s_is_jW_1^{s_i}W_1^{s_j}}\right>.\ee
Using the change of variable $w_i=\epsilon_i s_i$ and the property $\overline{\epsilon_i\epsilon_j}=0$ for $i\neq j$, as well as $G_1^{w_i}$ instead of $W_1^{s_i}$
(see Sec.~\ref{sec:time:deriv:mag}), we find
\be \overline{\langle \dot{m}^2\rangle}=\left\langle \frac{4}{N^2}\sum_{i=1}^N{G_1^{w_i}}^2\right\rangle.\ee
For $T>T_c$, we consider that $w, \dot{w}\ll 1$ so that we can write $\beta(J_1w+v)=w+\dot{w}$ at the lowest order in $w$ and $\dot{w}$. We have, 
\be {G_1^{w_i}}\sim 2^{-1}[1-w_i(w+\dot{w})]. 
\ee
Hence
\be \frac{4}{N}\left\langle \sum_{i=1}^N{G_1^{w_i}}^2\right\rangle \sim 1+\dot{w}^2-w^2\ee
and we finally find for $T\geq T_c$,
\be \overline{\langle \dot{m}^2\rangle}=N^{-1}\left(1-\langle w^2\rangle +\langle \dot{w}^2\rangle\right).\ee
For $T<T_c$, we have for any function $A(w_i)$ that
\be \label{eq:AA:appB}
\langle A(w_i)A(w_j)\rangle=\int_{-1}^1dmP(m)\sum_{\{w_i\}}P(\{w_i\}|m)A(w_i)A(w_j).
\ee
As the model expressed in terms of the variables $w_i$ and $v_i$ is a mean-field model, we have 
\be \label{eq:Pwm:appB}
P(\{w_i\}|m)=\left(\frac{1-m^2}{4}\right)^{N/2}\prod_{i=1}^N \left(\frac{1+m}{1-m}\right)^{w_i/2}.
\ee 
Using Eqs.~(\ref{eq:AA:appB}) and (\ref{eq:Pwm:appB}), 
we can show that for $T<T_c$,
\be\label{appendix:eq:av:A:A}\langle A(w_i)A(w_j)\rangle=\left\langle \left(\frac{(1+m)A(+1)+(1-m)A(-1)}{2}\right)^2\right\rangle.\ee
Thus, for $T<T_c$, one eventually finds
\begin{equation}\label{eq:mean:mdot2}
\overline{\langle \dot{m}^2\rangle}=N^{-1}\left(1-\langle w^2\rangle +\langle \dot{w}^2\rangle\right).
\end{equation}


\section{Computation of the non-linear susceptibility of the magnetisation}
\label{appendix:chi:nl:m:calcul}
Using the change of variable $w_i=\epsilon_i s_i$, as well as the property
\be \overline{\epsilon_i\epsilon_j\epsilon_k\epsilon_l}=(\delta_{i, j}\delta_{k, l}+\delta_{i, k}\delta_{j, l}+\delta_{i, l}\delta_{j, k}-2\delta_{i,j,k,l}),\ee
we find that
\be \overline{\langle m^4\rangle}=\frac{3}{N^2}-\frac{2}{N^3}
\ee
and
\begin{equation}\overline{\langle m^2\rangle^2}=\frac{1}{N^2}+\frac{2}{N^4}\sum_{i\neq j}\langle w_iw_j\rangle^2.\end{equation}
We thus obtain, 
\be \chi_{m}^{nl}=-2-\frac{6}{N}\sum_{i\neq j}\langle w_iw_j\rangle^2.\ee
Considering that all $\langle w_iw_j\rangle$ are equal for $i\neq j$, one also has
\be \langle w_iw_j\rangle=\frac{1}{(N-1)}(N\langle w^2\rangle -1),\ee
so that $\chi_{m}^{nl}$ becomes
\be  \chi_{m}^{nl}=-2-6\frac{N^2}{N-1}\left(\langle w^2\rangle-\frac{1}{N}\right)^2,\ee
which in the limit of large $N$ boils down to
\be \chi_{m}^{nl}\approx -2-6N\left(\langle w^2\rangle-N^{-1}\right)^2.\ee
At $T=T_c$, using the result of \cite{guislain_nonequil2023}, we have $P_N(w, \dot{w})=P_0\exp[-N b(v_0w^2+\dot{w}^2)^2]$ with $b$ given in \ref{appendix:coeff:lettre}.
We then find
\be \langle w^2\rangle= \gamma/\sqrt{N}\ee
with 
$\gamma=(v_0 \sqrt{\pi b})^{-1}$, which leads to the expression of $\chi_{m}^{nl}$ at $T=T_c$ given in Eq.~(\ref{eq:chim:nl:crit}).


\section{Computation of the non-linear susceptibility of the stochastic time derivative of the magnetisation}
\label{appendix:chi:nl:md:calcul}
Using the same method as for the magnetisation, we find
\be \label{appendix:eq:m4}\overline{\langle \dot{m}^4\rangle}=3\frac{2^4}{N^4}\sum_{ij}\langle  {G_1^{w_i}}^2{G_1^{w_j}}^2\rangle-2\frac{2^4}{N^4}\sum_{i=1}^N\langle {G_1^{w_i}}^4\rangle \ee
and 
\be\label{appendix:eq:m22} \overline{\langle \dot{m}^2\rangle^2}=\frac{2^4}{N^4}\left(\sum_{ij}\langle  {G_1^{w_i}}^2\rangle\langle {G_1^{w_j}}^2\rangle+2\sum_{i\neq j}\langle w_iw_j G_1^{w_i}G_1^{w_j}\rangle^2 \right).\ee
To lighten notations, we introduce the following quantities
\be \label{appendix:eq:G2G4} G_2=\frac{2^2}{N}\sum_{i=1}^N{G_1^{w_i}}^2 \quad \quad \text{ and }\quad \quad G_4=\frac{2^4}{N}\sum_{i=1}^N{G_1^{w_i}}^4.\ee
As, by definition of $\dot{w}$, we have \be\label{appendix:eq:3} \sum_{i,j} w_iw_j G_1^{w_i}G_1^{w_j} =\frac{N^2}{4} \dot{w}^2,\ee
we find that 
\be \sum_{i\neq j} 2^4\langle w_iw_j G_1^{w_i}G_1^{w_j}\rangle^2=\frac{N^3}{N-1}\left(\langle \dot{w}^2\rangle -\frac{1}{N}\langle G_2\rangle\right)^2.\ee
Injecting Eqs.~(\ref{appendix:eq:G2G4}) and (\ref{appendix:eq:3}) into Eqs.~(\ref{appendix:eq:m4}) and (\ref{appendix:eq:m22}), and then into the definition of the non-linear susceptibility of $\dot{m}$ [Eq.~(\ref{eq:chi:nl:mdot:def})], we find:
\be\label{appendix:eq:chi:nl:mdot} \chi_{\dot{m}}^{nl}\!=\!3 N\!\left(\!\left\langle G_2^2\right\rangle- \left\langle G_2\right\rangle ^2\!\right)-2\langle G_4\rangle -6N\!\left(\langle \dot{w}^2\rangle -N^{-1}\:\langle G_2\rangle\right)^2.\ee
For $T>T_c$, $\beta(J_1w+v)=w+\dot{w}$ at lowest order in $w$ and $\dot{w}$ thus 
\be G_1^{w_i}\sim\frac{1}{2}\big(1-w_i(w+\dot{w})\big).\ee
We find that the terms which contribute at this order are $G_2\sim 1+\dot{w}^2-w^2$, and $G_4\sim 1$, leading to
\be\label{appendix:eq:chi:nl:mdot:temp}
\chi_{\dot{m}}^{nl}=3N\left(\langle \dot{w}^4\rangle -3\langle \dot{w}^2\rangle^2+\langle w^4\rangle -\langle w^2\rangle^2 \right) -2.
\ee
For $T<T_c$, we find using Eq.~(\ref{appendix:eq:av:A:A}) that
\be G_2=1+\dot{w}^2-w^2\ee
and
\be G_4=\frac{1}{2}\left(-1+\dot{w}+w\right)^4\left(1-e^{8\tanh^{-1}(w+\dot{w})}(-1+w)+w\right).\ee
Injecting these two expressions into Eq.~(\ref{appendix:eq:chi:nl:mdot:temp}) and keeping the lowest order in $w$ and $\dot{w}$ gives Eq.~(\ref{appendix:eq:chi:nl:mdot}).
At $T=T_c$, using the result of \cite{guislain_nonequil2023}, we find 
\be \chi_{\dot{m}}^{nl}=-2+\rho,\ee
with $\rho$ given in Eq.~(\ref{eq:rho:value}).

\bigskip

\bibliographystyle{plain_url}
\bibliography{biblio}

\end{document}